\documentclass[conference]{IEEEtran}
\IEEEoverridecommandlockouts
\usepackage{cite}
\usepackage{amsmath,amssymb,amsfonts}
\usepackage{algorithmic}
\usepackage{graphicx}
\usepackage{textcomp}
\usepackage{xcolor}
\usepackage{float}
\usepackage{subfigure}
\usepackage{color}
\usepackage{transparent}
\usepackage{graphicx}
\usepackage{import}
\usepackage{hyperref}

\usepackage[misc]{ifsym}
\usepackage[ruled,linesnumbered]{algorithm2e}
\def\BibTeX{{\rm B\kern-.05em{\sc i\kern-.025em b}\kern-.08em
    T\kern-.1667em\lower.7ex\hbox{E}\kern-.125emX}}

\newcommand{\obsi}[1]{o_{{#1}}} % observation
\newcommand{\statei}[1]{s_{{#1}}} % observation
\newcommand{\hiddeni}[1]{h_{{#1}}} %
\newcommand{\acti}[1]{a_{{#1}}} %
\newcommand{\ret}[1]{R_{{#1}}} %
\newcommand{\rew}[1]{r_{{#1}}} %
\newcommand{\loc}[1]{x_{{#1}}} %

\usepackage{color} % TODO: delete laterp

\begin{document}

\title{SEA: A Spatially Explicit Architecture for Multi-Agent Reinforcement Learning}
% \author{\IEEEauthorblockN{Anonymous Authors}}
\author{\IEEEauthorblockN{Dapeng Li, Zhiwei Xu, Bin Zhang, Guoliang Fan
% \IEEEauthorrefmark{1}
% %\thanks{}
}
\IEEEauthorblockA{\textit{Institute of Automation, Chinese Academy of Sciences}}
\textit{School of Artificial Intelligence, University of Chinese Academy of Sciences}\\
Beijing, China\\
lidapeng2020@ia.ac.cn,xuzhiwei2019@ia.ac.cn, zhangbin2020@ia.ac.cn, guoliang.fan@ia.ac.cn
% \\
}

\maketitle

\begin{abstract}
% This document is a model and instructions for \LaTeX.
% This and the IEEEtran.cls file define the components of your paper [title, text, heads, etc.]. *CRITICAL: Do Not Use Symbols, Special Characters, Footnotes, 
% or Math in Paper Title or Abstract.
Spatial information is essential in various fields. How to explicitly model according to the spatial location of agents is also very important for the multi-agent problem, especially when the number of agents is changing and the scale is enormous. Inspired by the point cloud task in computer vision, we propose a spatial information extraction structure for multi-agent reinforcement learning in this paper. Agents can effectively share the neighborhood and global information through a spatially encoder-decoder structure. Our method follows the centralized training with decentralized execution (CTDE) paradigm. In addition, our structure can be applied to various existing mainstream reinforcement learning algorithms with minor modifications and can deal with the problem with a variable number of agents. The experiments in several multi-agent scenarios show that the existing methods can get convincing results by adding our spatially explicit architecture.
\end{abstract}

% \begin{IEEEkeywords}
% % component, formatting, style, styling, insert
% Multi-Agent System, Spatial Information, Coordination and Collaboration
% \end{IEEEkeywords}

\section{Introduction}
In recent years, multi-agent reinforcement learning (MARL) has made lots of progress~\cite{dota2drl,starcraft2,trafic_chu,autonomous_driving}. Despite the success of the above achievement, few MARL algorithms have been successfully applied to real-world applications. One important reason is that the problems will become incredibly complex due to the enormous number of agents in some environments. If we simply use the independent single-agent RL algorithm, it is often difficult to form cooperation without considering the information of other agents~\cite{independent_learning}. However, using centralized methods to control all agents will lead to an exponential explosion of state and action space~\cite{Busoniu2010}. In addition, the number of agents in many scenarios will constantly change between and within episodes. Therefore, two main concerns can be summarized: a) How to effectively use or combine global and local information. b) How to avoid the curse of dimensionality when the scale of agents increases.
% The model structure is not affected by the number of agents.

To address the above issues, several studies~\cite{mfrl,ncc,mfvfd,DGN} try to simplify the problems by only considering the information of neighboring agents rather than all other agents. However, this proximity approximation will lose global information and may lead to the loss of global cooperation. Besides, most of them simply calculate the adjacency matrix, which ignores the variation in the influence of neighbors at different distances.

Some approaches use specific network structures to share information between agents during execution or training. For example, CommNet~\cite{commNet} uses average pooling for the information of all agents to promote communication between agents. However, the simple average operation will lose specific information about each agent. Therefore, BicNet~\cite{bicnet} utilizes the recurrent neural network (RNN) to pass information between agents to avoid the above defect. However, the input order of agents will affect the outputs of RNN, so such methods will become unstable when the order and number of agents are not fixed. 
% In addition, ATOC~\cite{atoc}, TarMAC~\cite{tarmac}, and MAAC~\cite{maac} use the attention mechanism to achieve information sharing among all agents.

Another branch in MARL is the grid-wise method~\cite{gridnet}. Similar to the segmentation task in computer vision, GridNet~\cite{gridnet} divides the continuous environments into discrete grid-wise space and gives the prediction to each grid. By feeding the global state into convolution networks in the encoder-decoder architecture, the joint spatial representation of all agents can be abstracted. Nevertheless, this grid-wise structure greatly limits its accuracy and practicability.

The above examples show that using spatial location and information sharing among agents are very meaningful for solving multi-agent problems, but each of them faces unique challenges. Therefore, designing a general framework that can take into account the spatial information of agents and effectively extract the information between agents is a precious problem. We discovered a significant similarity between the point cloud classification problem and the large-scale multi-agent problem. Specifically, both have some points (or agents) with spatial topology, and each point contains multiple features. The prediction of each point is based on the information of other local points or all points. Inspired by a series of classical works~\cite{pointnet,pointnet2} in the point cloud problem, we propose a general \textbf{S}patially \textbf{E}xplicit \textbf{A}rchitecture~(\textbf{SEA}), which treats each agent as a coordinate point with features and learns a joint spatial representation through an encoder-decoder structure. The agent set is first divided into overlapping local clusters by the encoder according to the locations of agents. Then the local features are extracted from each small cluster. Through multi-level extraction, the receptive fields of agents are continuously expanded. So the clusters at the higher level contain global features. In the decoder, the features of each cluster are reconstructed according to its location. Therefore, the output features contain both local and global information. 
The advantages of SEA can be summarized as follows: 
\begin{itemize}
\item SEA can promote local and global information sharing and effectively utilize spatial information.

\item SEA can deal with large-scale and variant number agent problems.

\item The output result of SEA is not affected by the input order of agents.
\end{itemize}

In this paper, we combine SEA with the current mainstream MARL algorithms. The experimental results in continuous and discrete observation scenarios demonstrate that the current mainstream reinforcement learning algorithms can be significantly improved by introducing SEA.

\section{Related Work}
Our paper is mainly related to two strands of academic literature: MARL and spatially explicit learning. 

\subsection{Multi-agent Reinforcement Learning}
In the MARL problem, the simplest training method is to train and execute each agent independently and treat other agents as part of the environment~\cite{independent_learning}. Although this fully decentralized method is easy to be implemented and has lower computation cost, it is unstable because each agent does not consider the dynamic of other agents. The inverse method, also known as the fully centralized method, uses a centralized model to make decisions for all agents based on global information. However, the state space and decision space will grow exponentially with the increase in the number of agents. 
% It is difficult for each agent to obtain the information of all other agents in the actual implementation. 
For the centralized method, the existing research primarily designs a specific strategy structure, including using RNN and attention models to achieve information sharing between agents or directly inputting the global state to get the actions of each agent through multiple heads.

In addition to the above two extreme methods, another compromise is Centralized Training with Decentralized Execution (CTDE). In the CTDE setting, agents can share information during training while making decisions based on only local observations during execution. This approach has received extensive research, primarily involving centralized policy gradient methods and value decomposition methods. Each agent in the centralized policy gradient method consists of a centralized critic and a decentralized actor. The centralized critic has access to the global information of all agents. The classic centralized policy gradient works include MADDPG~\cite{maddpg}, COMA~\cite{coma}, and MAPPO~\cite{mappo,efficient}. Value function decomposition methods decompose the joint state-action value function into individual ones. For example, VDN~\cite{vdn} represents the joint Q-value as the sum of individual ones, and QMIX~\cite{qmix} uses a parameterized mixing network to calculate the joint Q-value. Due to the excellent performance of QMIX, a series of variants~\cite{side,haven,mbvd} have been proposed to solve various problems.
 
\subsection{Spatially Explicit Learning}
Spatial information has been widely used in various fields, such as social networks and ecology. Many studies have shown that models that explicitly consider spatial relationships will outperform those that do not~\cite{serl}. In the multi-agent field, there are also many works~\cite{mfrl,ncc,DGN,pool} to achieve local cooperation by considering the information of neighbor agents. They show that models can perform better by introducing spatial information. Due to only considering the neighborhoods, most such methods can handle situations with a varying number of agents. However, the above methods only concern local information but ignore global sharing.

We are inspired by other research fields considering spatial information, like the point cloud segmentation problem in computer vision. The most challenging problem in point cloud segmentation is handling such a tremendous and variant number of spatial points with features, which also exists in the MARL situation. Moreover, trade-offs for the global and local features are required in both the point cloud segmentation and the multi-agent problems. For point cloud segmentation problems, PointNet~\cite{pointnet} uses simple max pooling that only obtains global information. Its advanced work, PointNet++~\cite{pointnet2}, tries to use multi-level structures to treat such a trade-off.

\section{Background}

\subsection{Problem Formulation}

Reinforcement learning can be viewed as the process where agents learn to obtain the maximum return by sequentially interacting with the environment. The cooperative multi-agent problem in this paper can be formulated as a tuple $G=\left< S,A,N,O,\mathcal{T}, Z, \rew{i},\gamma\right>$. $\statei{}\in S$ is the true global state of the environment. At each discrete time step $t$, each agent $i\in\mathcal{N}:=\{1,\dots,N\}$ will select an action $\acti{i}\in A_i$, where $A=A_1\times\cdots\times A_N$ is the joint action space. $\mathcal{T}(s'|s,\boldsymbol{a}):S\times A \times S \rightarrow P(S)$ is the state transition function, where $\boldsymbol{a}=\{a_1,\dots,a_N\}$ is the joint action. In the partially observable Markov decision process (POMDP), the global state of the environment is not accessible and each agent $i$ can only get its individual observation $\obsi{i}\in O$ by the observation function $Z(s, i): S \times \mathcal{N} \rightarrow O$. $r_i(s,a_i):S\times A_i \rightarrow \mathbb{R}$ is the reward function for each agent $i$.  The return $\ret{i}=\sum^T_{t= 0}\gamma^tr^{t}_i(s,a_i)$ is defined as the discounted sum of rewards, where $\gamma\in [0,1]$ is the discount factor. The goal of each agent $i$ is to learn the policy $\pi_i(a_i|\obsi{i}):O_i \rightarrow P(A_i)$ to maximize its expected return.

% ~\cite{decpomdp}
\subsection{Centralized Policy Gradient Methods}
In single-agent tasks, value-based RL methods often need to evaluate the optimal cumulative reward and determine the optimal policy based on it. In policy-based methods, however, the policy function is directly optimized, and the evaluation of the value function is optional and secondary. Policy gradient methods always consist of a parameterized policy function, often called the "actor." The actor's parameters can be directly optimized by gradient descent to find the optimal policy. The policy gradient can be computed by the following equation:
\begin{equation} 
    g =\mathbb{E}\left[\sum_{t=0}^{\infty} \Psi_t \nabla_\theta \log \pi_\theta\left(a_t \mid s_t\right)\right],
\end{equation}
where $\Psi_t$ can be represented differently by various algorithms. 
% In order to assist policy optimization and speed up training, some algorithms also introduce the ``critic" to evaluate the value function. 
In addition to the early work represented by Advantage Actor Critic (A2C), policy gradients have two branches known as trust region-based methods and deterministic policy gradient methods. 
% The policy gradient methods described above make significant contributions to solving single-agent reinforcement learning tasks.
Compared with the fully decentralized and fully centralized methods, CTDE has gradually become a more popular multi-agent training paradigm as a compromise method. In the centralized training phase, agents can obtain information such as the observations or policies of all other agents but can only select appropriate actions according to their own local observations during decentralized execution. Therefore, the CTDE-based method can alleviate the instability of the environment, and it is also qualified for partially observable multi-agent tasks. The centralized policy gradient methods, including MAA2C, MADDPG, and MAPPO, apply the single-agent policy gradient approach to the multi-agent problems and follow the CTDE paradigm. Taking advantage of the Actor Critic framework, such a method can use the actor model to make decisions at execution time and evaluate the joint actions of all agents through the critic when training. 

\begin{figure*}[ht]
    \centering
    \includegraphics[width=6.2 in]{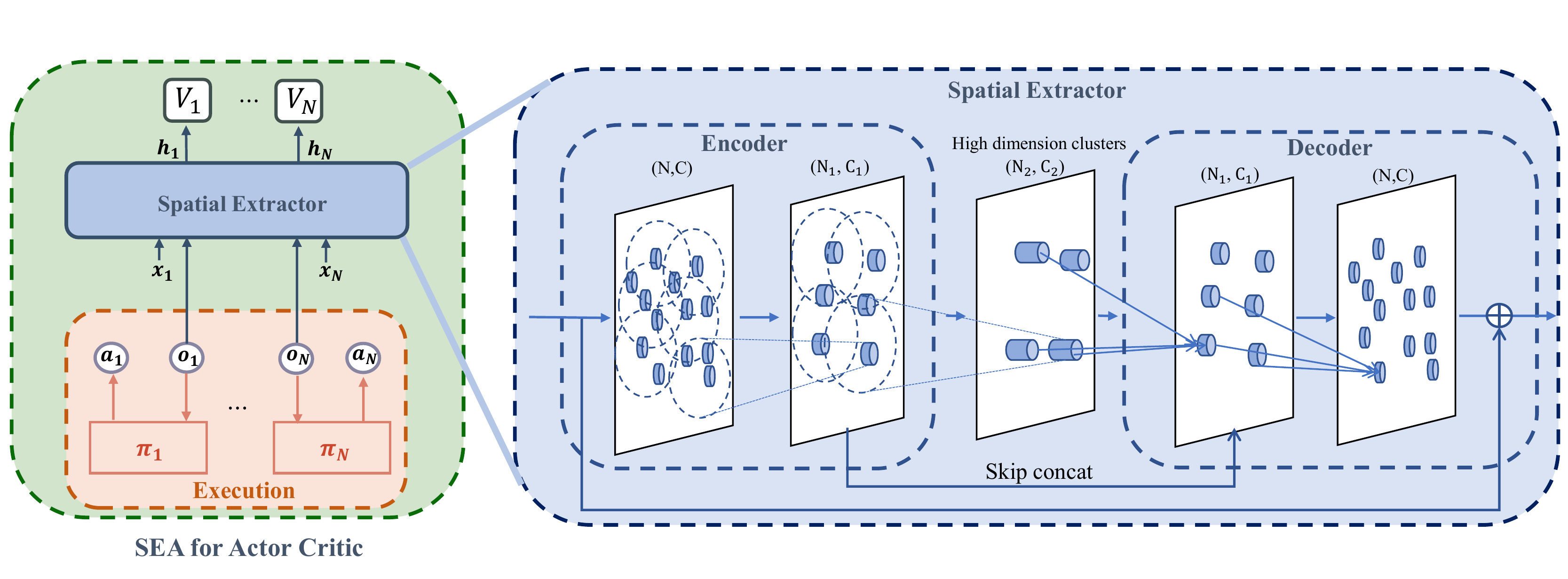}
    \caption{The overview design of SEA-AC. The left part is the schematic of SEA in Actor Critic structure, and the right part is the framework of SEA structure.}
    \label{fig:overview}
\end{figure*}

The CTDE methods have recently progressed in the multi-agent field. However, such methods do not consider the spatial relationship between agents but take position information as a regular feature and pass it into the model for learning without distinguishing it from other features. These implicit methods disregard much spatial information and the relationship between agents.
\newline
\section{Methodology}
\subsection{Spatial Representation from Point Wise View}
In this chapter, we first briefly introduce the commonness between point cloud segmentation and multi-agent cooperation, then build a connection between the point cloud and multi-agent problem by considering the agent from a point wise view. Therefore, we can naturally use many structures in the point cloud domain to solve the spatial representation problem in multi-agent problems.

The purpose of point cloud segmentation is to classify massive point sets. For unordered point sets $\{\loc{1},\loc{2},\dots,\loc{N}\}$ with $\loc{i}\in\mathbb{R}^d$, PointNet uses max pooling operation to aggregate the whole point set to a vector:
\begin{equation}
f(\loc{1},\loc{2},\dots,\loc{N})=\phi({\rm MAX_{i=1,\dots,N}}\{\beta(\loc{i})\}),
\label{pointnet}
\end{equation}
where $\phi$ and $\beta$ are MLP networks.
Although this simple operation has some good properties, it loses local feature information of different scales. Therefore, a multi-scale feature representation is proposed in PointNet++, which can extract global and local information with different scales.

Similarly, in most multi-agent environments, agents have their geographic location \cite{magent}. Therefore, building a spatial model in multi-agent problems to extract the spatial relationships between agents is a practical problem. However, most algorithms ignore this information and treat location as a regular feature which will cause a loss of much important information. In such a scenario, we hope that the model can have the following good properties: (a) The model is invariant to input permutation; (b) the model structure can handle a variable number of agents; (c) the model can extract both global and local information. 
Therefore, this problem is very similar to the point cloud segmentation problem. Specifically, both of them have a large number of points with spatial topology, and each point has its own features. The prediction of each point needs to consider both the neighborhood and global points. Therefore, we can regard each agent as a feature point with spatial coordinates. The coordinate position of all feature agents is $\{\loc{1}, \dots, \loc{N}\}$, and the local observation feature of agents is $\{
\obsi{1}, \dots, \obsi{N}\}$

Inspired by PointNet++, we use an encoder-decoder structure as the spatial extractor. This structure generates high-dimensional cluster center features points with different vision range through multi-level down-sampling and then interpolate new features with local and global information for each agent through multi-level up-sampling.
In the encoder, we first use farthest point sampling \cite{pointnet2} to choose $N_l$ agents as the centroids of clusters for grouping at layer $l$, where $l\in\{1,2\}$ is the number of down sample layers. $N\ge N_1 \ge N_2$, and $N$ is the number of input agent. Then divide the nearest $K$ agents for each cluster centroid as its neighborhood. Note that each agent can be divided into multiple clusters. 
  % The detail description of FPS can be found in Appendix.

% unlike CommNet and mean-filed only consider the average of local or global information extremely, we can both extract the global and local features by this multi-level structures.
For each cluster, we can use PointNet (as is Eq.\ref{pointnet}) to extract features in each cluster. For a points subset coordinates$\{\loc{1},\dots,\loc{n}\}$ and its local observation $\{\obsi{1},\dots,\obsi{n}\}$ in one cluster, where $n$ is the number of agents in this cluster, the common features can be extracted as follows:
\begin{equation}
f(\obsi{1},\dots,\obsi{n})=\phi({\rm MAX_{i=1,\dots,n}}\{\beta(\obsi{i})\}).
\label{max}
\end{equation}
% we give two choices to extract the grouping features. The first is using average pooling to extract common local features.
% \begin{equation}
% f(\obsi{1},\dots,\obsi{n})=\gamma({\rm MEAN_{i=1,\dots,n}}\{\beta(\obsi{i})\}).
% \label{avg}
% \end{equation}

Suppose $x_j$ is the center point of this cluster; the output of this cluster $f_j$ and the coordinates of the center point $\loc{j}$ will consist of a new point at a higher level. The reason for using MLP layer $\beta$ and $\phi$ is that the entire cluster has more information than a single point, so we need to use a layer to increase the shape of the feature vector to avoid losing too much information in the grouping process. By repeating this up-sampling process, we can get a multi level point subset, and hence the points at a higher level have a bigger vision range. Since the max operation is a symmetry function, the output is invariant to the input agent's order. 

In the decoder, calculate the inverse distance between agents in layer $l-1$ and its nearest $k$ agents in layer $l$ as the interpolation weight, and generate new features of $N_{l-1}$ agents in layer $l-1$ by interpolating. The interpolating process can be formed as follows:
\begin{equation}
h_{j}=\frac{\sum^k_{i=1}w_i(x_j)f_{i}}{\sum^k_{i=1}w_i(x_j)}\quad w_i(x)=\frac{1}{d(x,x_i)^p},
\end{equation}
where $d$ is the distance metric.

\subsection{Spatially Explicit Architecture for MARL}
Unlike the end-to-end application of this structure to segmentation in computer vision, we use this encoder-decoder structure as an intermediate plugin to extract spatial representation. Therefore, this structure's output and input dimensions are set to the same, which is more versatile and easy to migrate to different MARL algorithms. We also add an additional residual structure. In addition, we remove the Batch Norm layer because of the instability of using the Batch Norm layer in reinforcement learning. The entire architecture is shown on the right side of Fig.\ref{fig:overview}.

Next, we will describe how to combine this spatially explicit architecture with the existing multi-agent frameworks, including A2C, PPO, and DDPG.

\paragraph{SEA-A2C} 
In order to meet the setting of CTDE, we do not use additional information in the actor. As shown on the left side of Fig.\ref{fig:overview}, in critic, by setting the coordinate $\loc{i}$ of agent and its corresponding observation $\obsi{i}$ constructed as a feature point $\{x_i, \obsi{i}\}$. Input all feature points in the same frame into Spatial Extractor, and each agent can obtain its new hidden features with local and global information,
\begin{equation}
\boldsymbol{\hiddeni{}}=\{\hiddeni{1},\dots,\hiddeni{N}\} ={\rm SE}(x_1,\dots,x_N,o_1,\dots,o_N),
\end{equation}
then input the new hidden feature $\hiddeni{i}$ of each agent into MLP to obtain the current state value of each agent. We call this structure SEA-A2C. 

\paragraph{SEA-PPO}
% For scenarios with fixed number of agents, w
The PPO algorithm is similar to A2C since it also computes a state value function, but it has a gradient clipping and can train multiple epochs for the same trajectories. Since we only change the input of the critic, the loss function of the critic network can be formed as follows:
\begin{equation}
\begin{split}
\begin{aligned}
\mathcal{L}(\theta)=\sum^N_{i=1}(\max[(V_\theta(\hiddeni{i})-\hat{R})^2,({\rm clip}(V_\theta(\hiddeni{i})\\ ,V_\theta(\hiddeni{i})-\varepsilon,V_\theta(\hiddeni{i})+\varepsilon)-\bar{R})^2]).
\end{aligned}
\end{split}
\end{equation}

\begin{figure*}[ht]
\subfigure[]{
\begin{minipage}{0.32\linewidth}
\centerline{\includegraphics[width=1.1\textwidth]{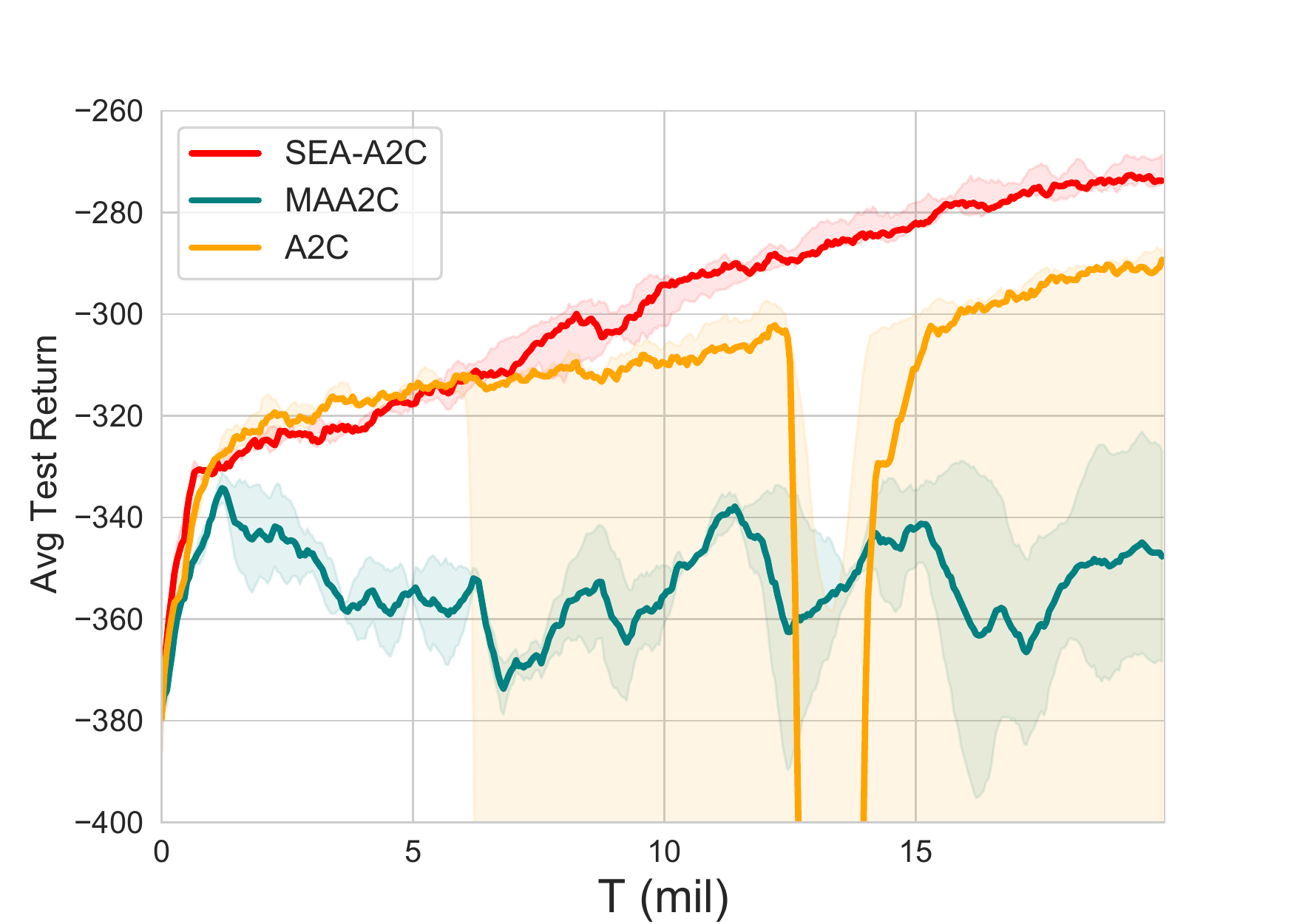}}
\label{a2c_}
% \centerline{(a)}
\end{minipage}}
\subfigure[]{
\begin{minipage}{0.32\linewidth}
\centerline{\includegraphics[width=1.1\textwidth]{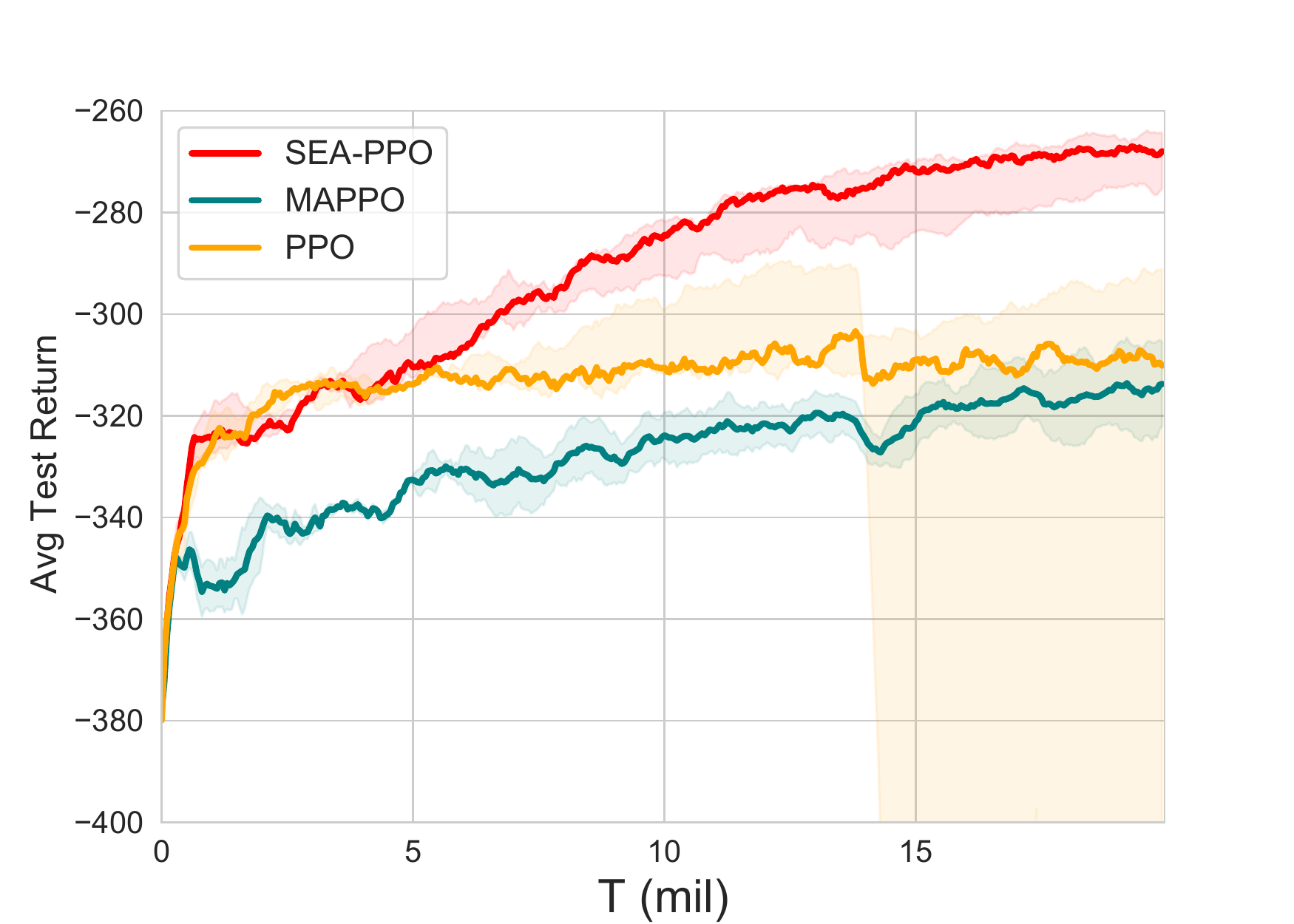}}
% \centerline{(b)}
\label{ppo_}
\end{minipage}}
\subfigure[]{
\begin{minipage}{0.32\linewidth}
\centerline{\includegraphics[width=1.1\textwidth]{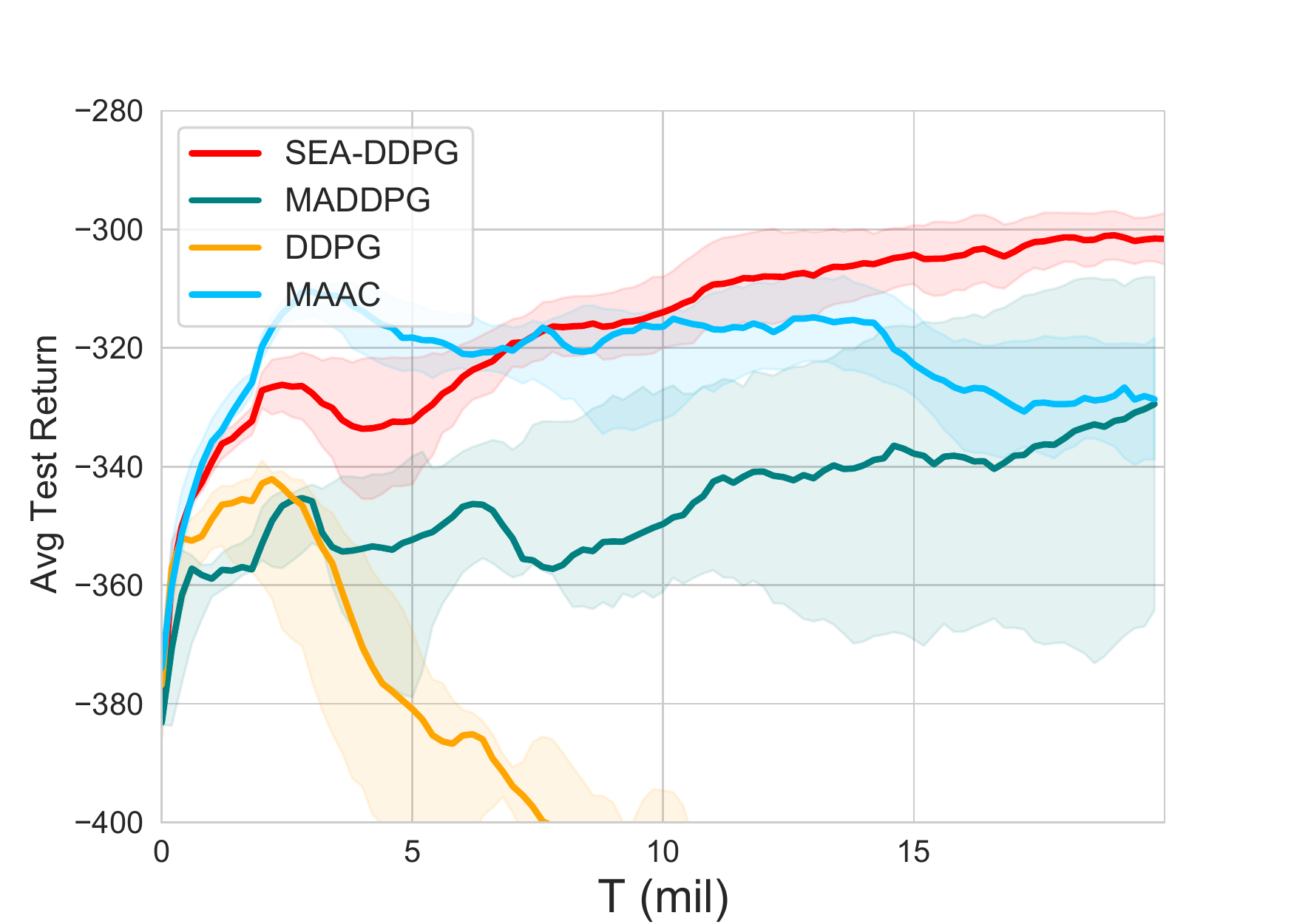}}
% \centerline{(c)}
\label{ddpg_}
\end{minipage}}

\caption{The agent average test return of different algorithms on \emph{Cooperative Navigation}.}
\label{fig:mpe_curve}
\end{figure*}
% \begin{figure*}[ht]
% 	\centering
% 	\subfloat[\label{fig:a}]{
% 		\includegraphics[scale=0.33]{figure/mpe_SimpleSpread-6ag-v0_test_return_a2c.pdf}}
% 	\subfloat[\label{fig:b}]{
% 		\includegraphics[scale=0.33]{figure/mpe_SimpleSpread-6ag-v0_test_return_ppo.pdf}}
% 	\subfloat[\label{fig:c}]{
% 		\includegraphics[scale=0.33]{figure/mpe_SimpleSpread-6ag-v0_test_return_ddpg.pdf}}
% 	\caption{Test return of different methods on Cooperative Navigation of different algorithms.}
% 	\label{mpe} 
% \end{figure*}

\paragraph{SEA-DDPG}
While the SEA-DDPG learns an action-value function since the input is observation and action, so the hidden state of each agent can be formed as follows:

\begin{equation}
% \begin{split}
\boldsymbol{\hiddeni{}}
% &=\{\hiddeni{1},\dots,\hiddeni{N}\} \\
={\rm SE}(\loc{1},\dots,\loc{N},\obsi{1},\dots,\obsi{N},\acti{1},\dots,\acti{N}),
% \end{split}
\end{equation}
and the policy gradient can be calculated as:
\begin{equation}
\begin{aligned}
\nabla J(\pi_i)=\mathbb{E}_\mathcal{D}[\nabla{\pi_i}(\tau_i)\nabla_{a_i}Q_i(\hiddeni{i})|_{\acti{i}=\pi(\tau_i)}],
\end{aligned}
\end{equation}
$Q_i(h_i)$ represent the action-value function.
 
The advantage of SEA is that it can not only realize the feature representation based on spatial location between agents but also has the ability to handle the problems under the varying or large scale of agent number.

\paragraph{Cluster setting and Death Masking}
Two important properties in many multi-agent scenarios are the number of agents that keep changing during one episode(most of them keep decreasing due to the agent's demise) or the scale of agents in different scenarios varies greatly. Such a variant will cause two main problems for our algorithm: how to deal with those dead agents in training and how to set a suitable number of clusters in each layer for different scale tasks. For the first question, we only need to filter out those dead agents at the first layer in the encoder, otherwise, those dead agents will also be divided into the cluster and affect the results. To solve the second question, we simply set the size of the cluster to be proportional to the size of the agent in each scenario. For a scenario with $n$ agents, then the number of clusters in layers 1 and 2 can be calculated as follows:
\begin{equation}
\begin{aligned}
N_1 &= \lfloor\sqrt{n}\rfloor\\
N_2 &= \lfloor\sqrt{N_1}\rfloor.
\end{aligned} 
\end{equation}
\newline

\begin{figure}[ht]
\begin{minipage}{0.48\linewidth}
\centerline{\includegraphics[width=1.3\textwidth]{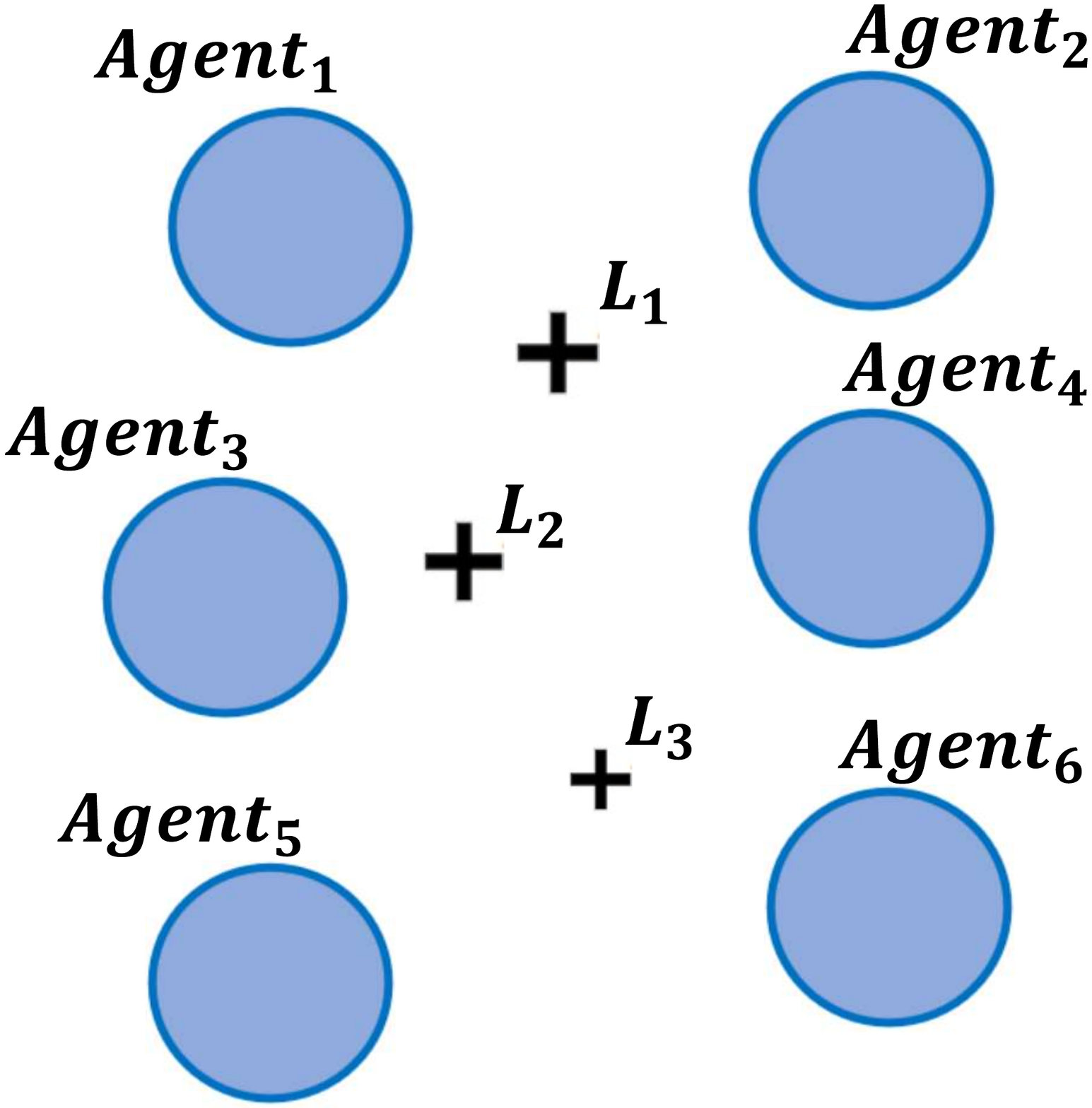}}
\centerline{(a)Cooperative Navigation}
\end{minipage}
\begin{minipage}{0.48\linewidth}
\centerline{\includegraphics[width=1.3\textwidth]{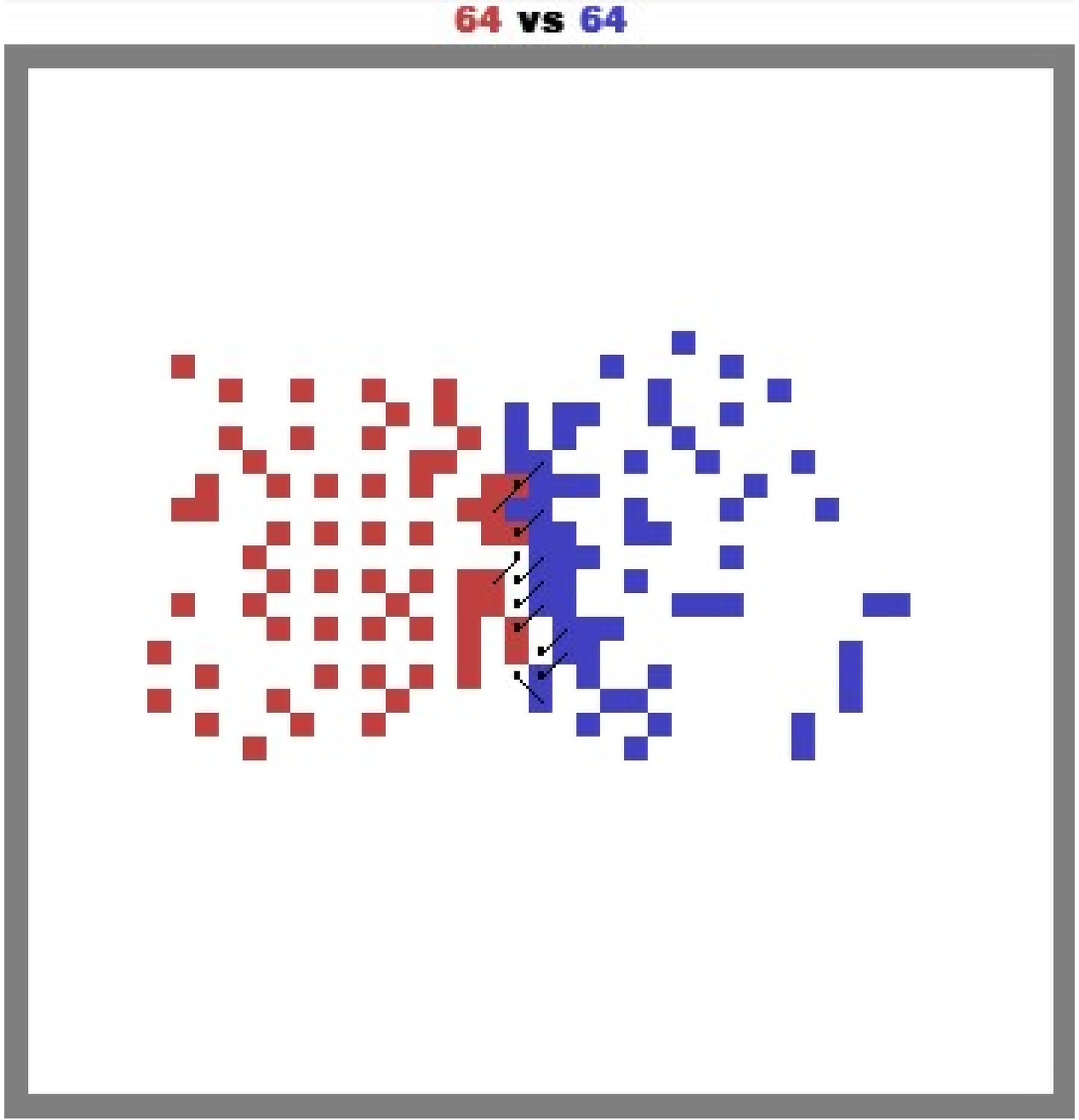}}
\centerline{(b)Battle}
\end{minipage}

\begin{minipage}{0.48\linewidth}
\centerline{\includegraphics[width=0.9\textwidth]{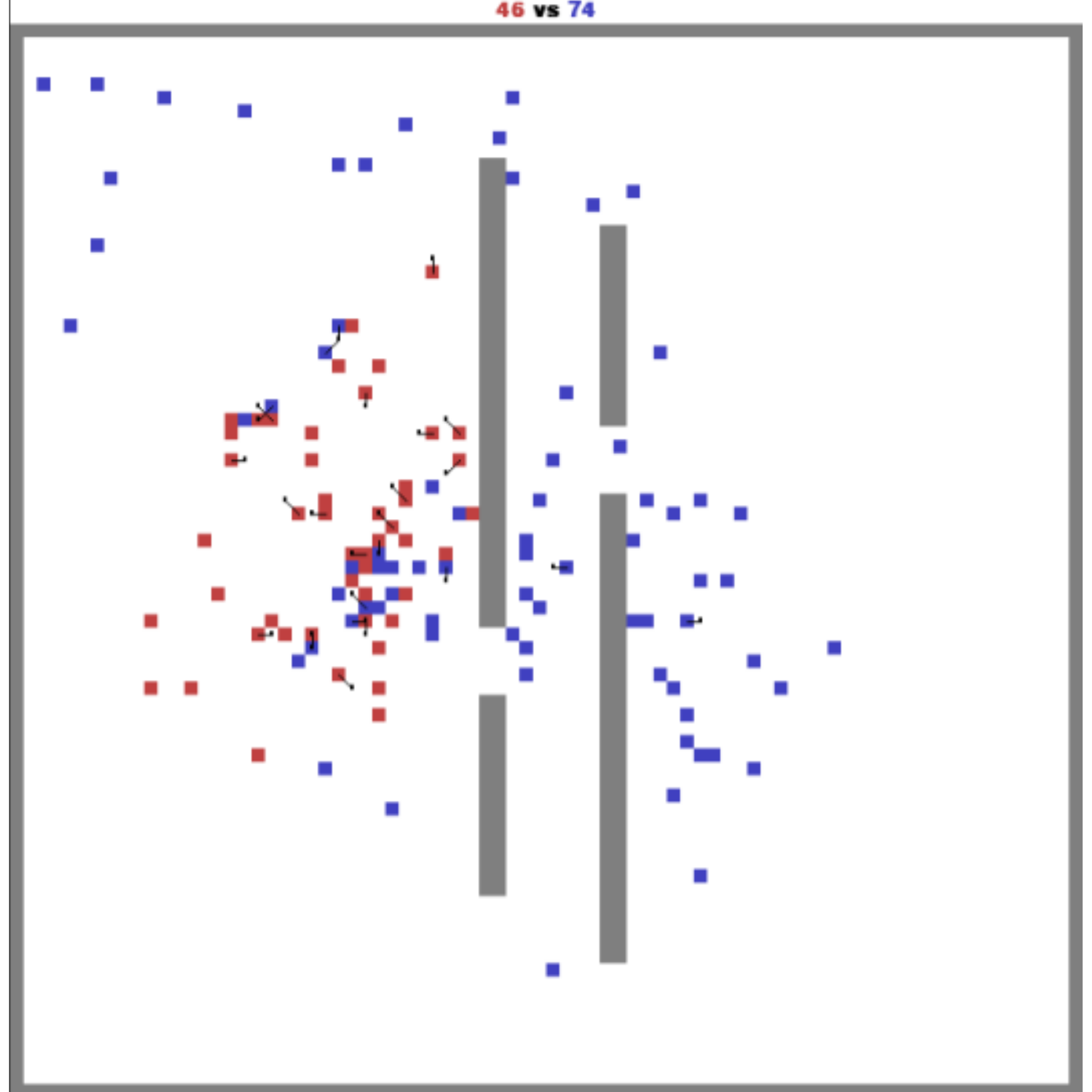}}
\small{\centerline{(c)Battlefiled}}
\end{minipage}  
\begin{minipage}{0.48\linewidth}
\centerline{\includegraphics[width=1.3\textwidth]{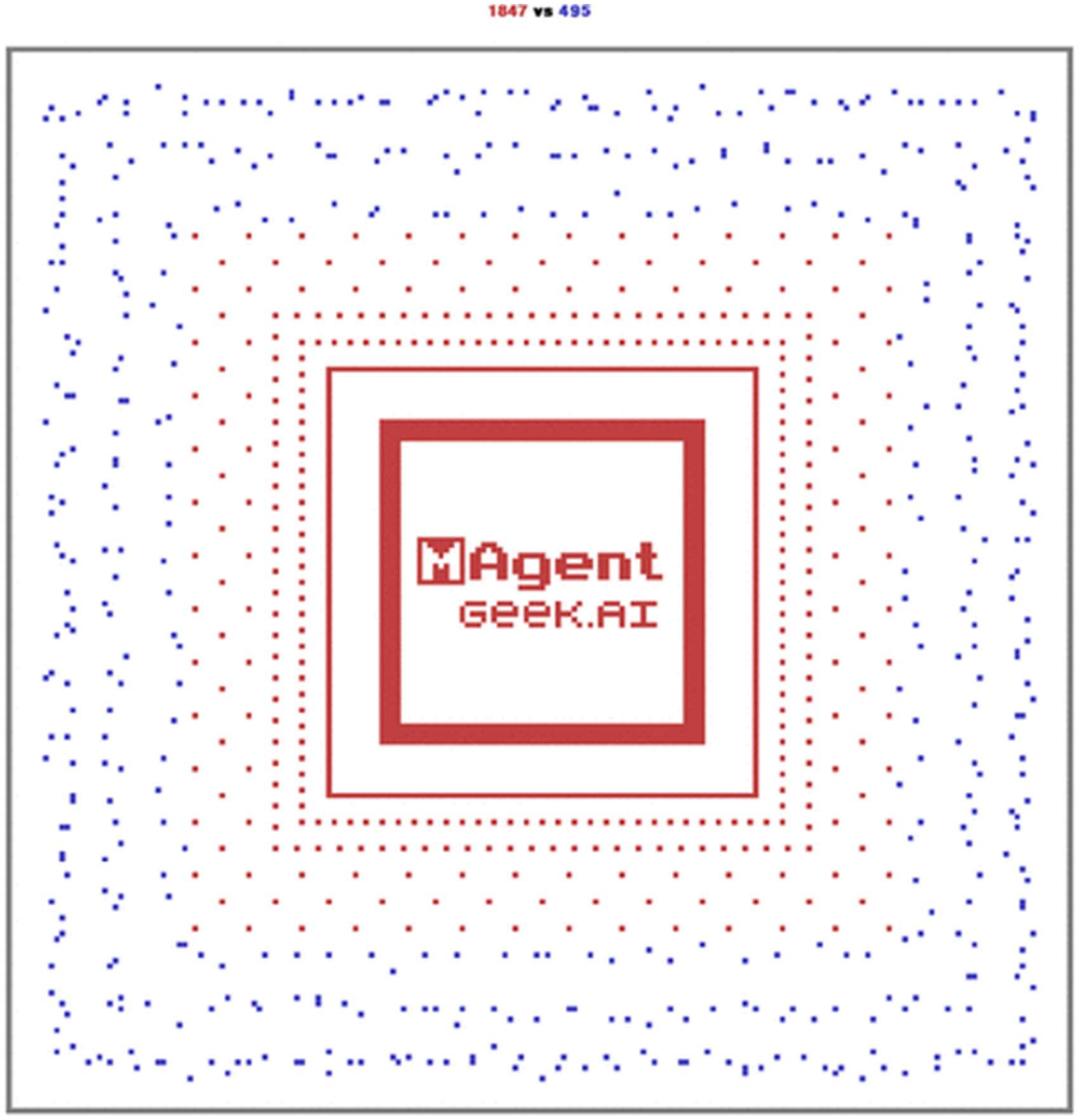}}
\centerline{(d)Gather}
\end{minipage}
\caption{The visualization of the four environments that we carry out our experiments on. (a)\emph{Cooperative Navigation}(b)\emph{Battle} (c)\emph{Battlefiled} (d)\emph{Gather}.}
\label{envs}
\end{figure}

\section{Experiments}
To evaluate the ability of SEA in different agent scales, we evaluate SEA in four multi-agent environments including \emph{Cooperative Navigation}~\cite{lowe2017multi}, \emph{Battle}, \emph{Battlefield} and \emph{Gather}~\cite{magent}. The number of agents in these scenarios ranges from a few to several hundred and includes both discrete observation and continuous observation scenes. 
% The visualization and detailed information of all environments can be found in Appendix A.

Through these experiments, we would like to answer the following questions:

\textbf{Q1}: How much improvement can the SEA method bring to the existing methods?

% \textbf{Q2}:How about the performance of SEA in scenarios with varying number of agents?

% \textbf{Q3}:How about the generalization of SEA model in different number of agent scenarios?

\textbf{Q2}: How does it compare with other algorithms also interact information between agents (e.g Attention)?

\textbf{Q3}: How do local and global information affect the performance of SEA?

\begin{figure*}[ht]
\subfigure[Battle]{
\begin{minipage}{0.32\linewidth}
\centerline{\includegraphics[width=1.1\textwidth]{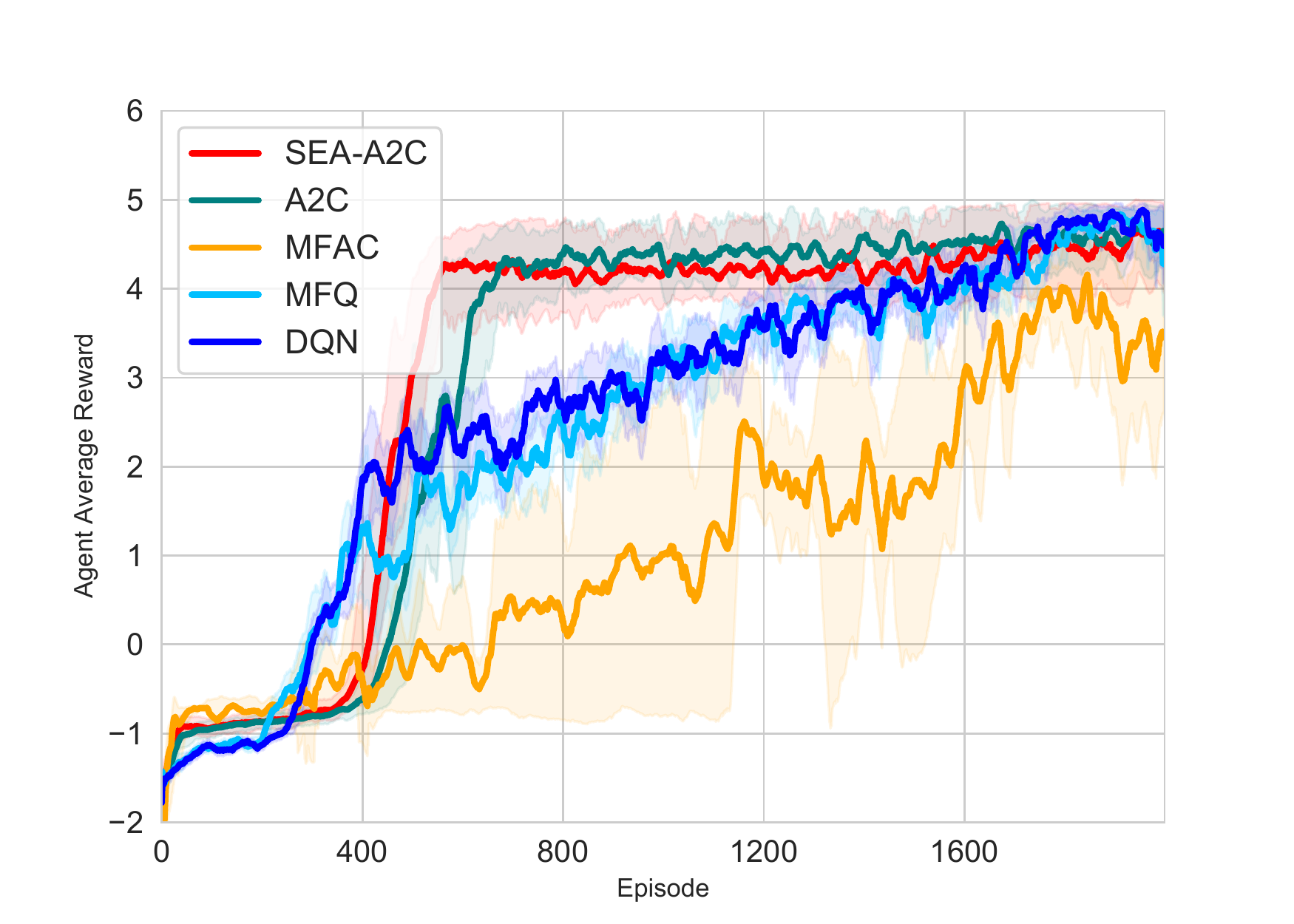}}
\label{fig:battle}
% \centerline{(a)}
\end{minipage}}
\subfigure[Battlefield]{
\begin{minipage}{0.32\linewidth}
\centerline{\includegraphics[width=1.1\textwidth]{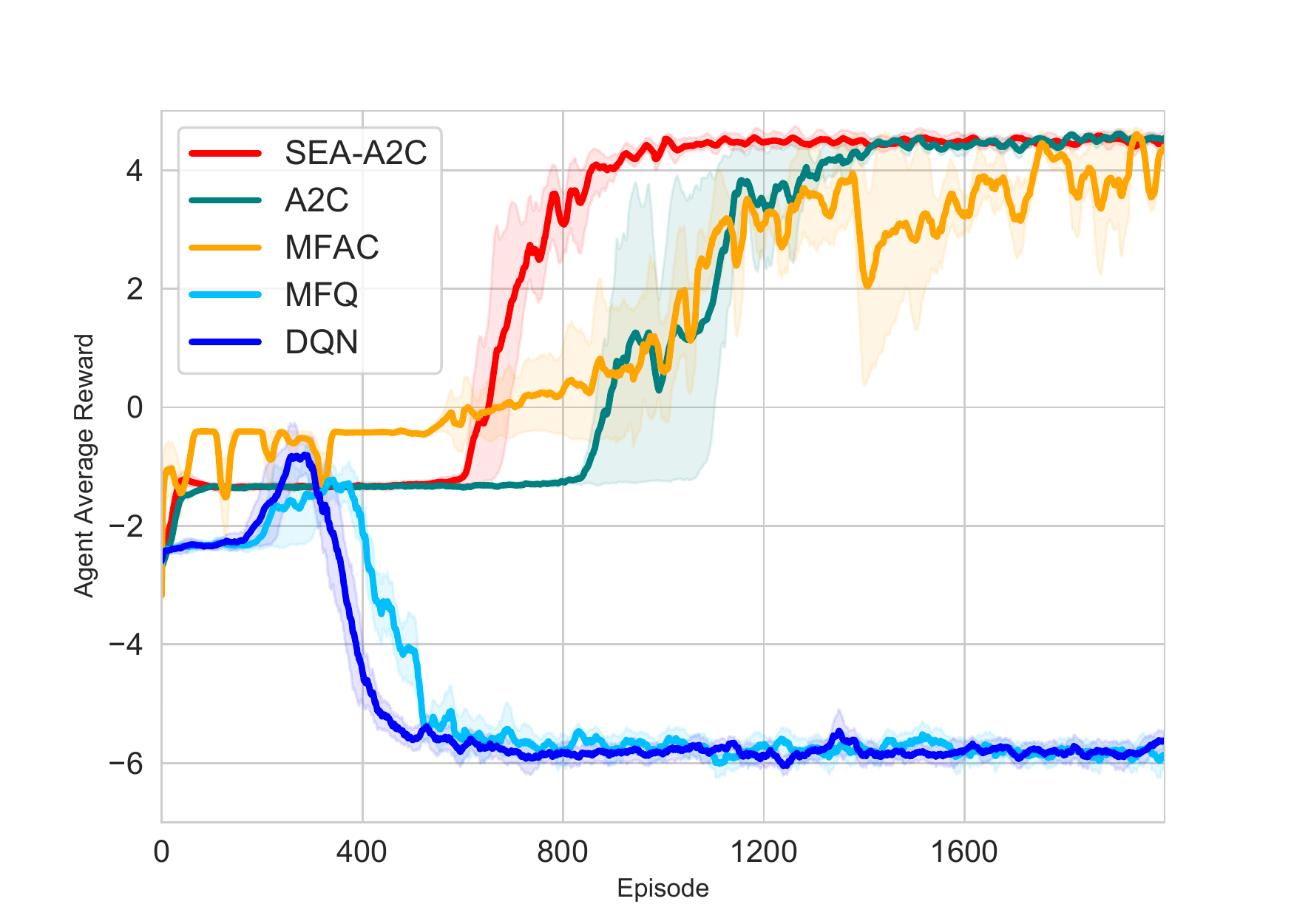}}
% \centerline{(b)}
\label{fig:battlefiled}
\end{minipage}}
\subfigure[Gather]{
\begin{minipage}{0.32\linewidth}
\centerline{\includegraphics[width=1.1\textwidth]{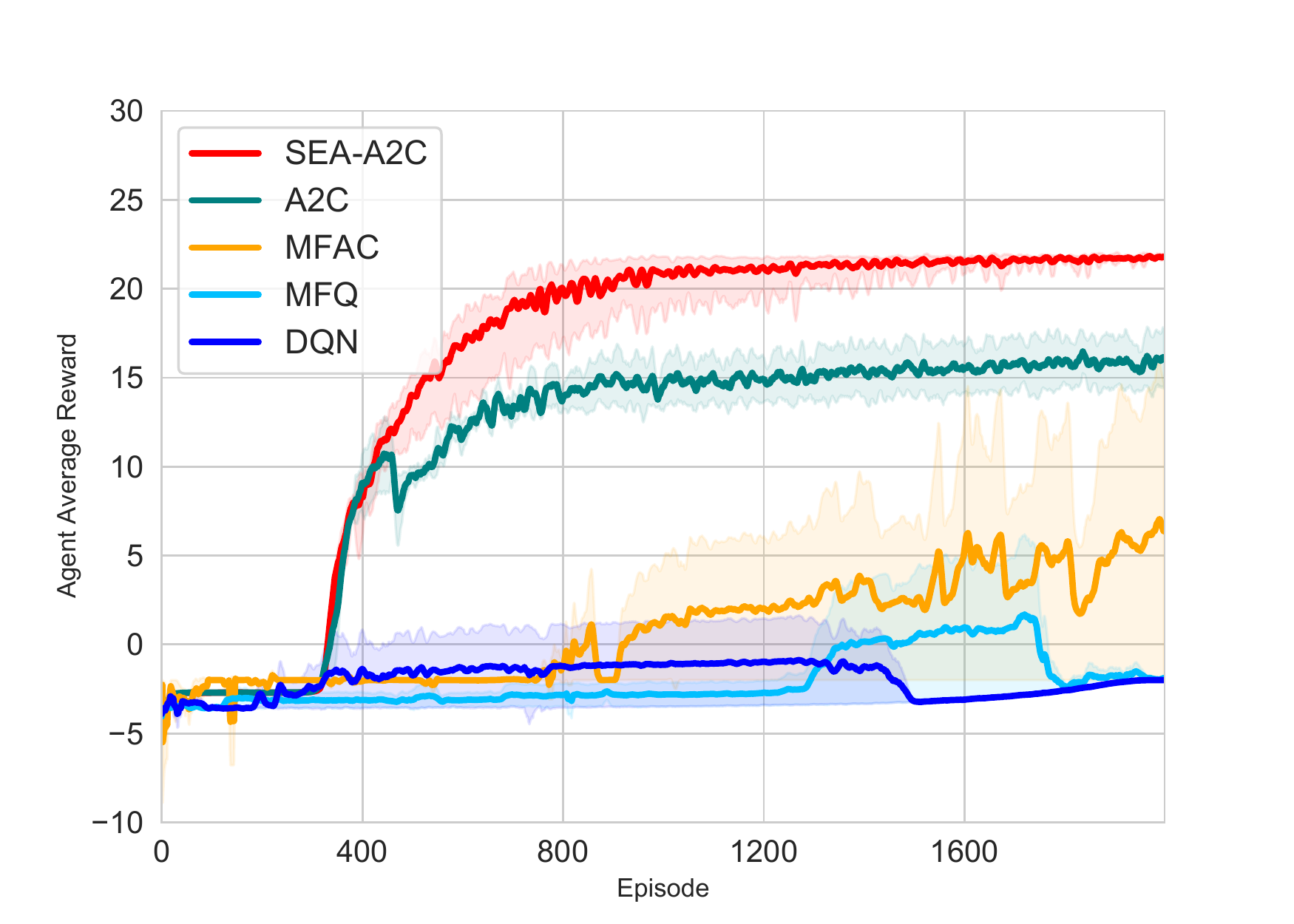}}
% \centerline{(c)}
\label{fig:gather}
\end{minipage}}

\caption{The agent average test return of different algorithms on \emph{Battle}, \emph{Battle Filed} and \emph{Gather}.}
\label{fig:magent_curve}
\end{figure*}

\subsection{Environment Description}

We tested our algorithm on four multi-agent environments including \emph{Cooperative Navigation}, \emph{Battle}, \emph{Battlefiled}, \emph{Gather}. 
The visualization of all environments can be found in Fig.\ref{envs}, and each environment we tested is described in detail below.

\noindent\textbf{Cooperative  Navigation} The \emph{Cooperative Navigation} is a task from Multi-Agent Particle Environment. This task involves particles and landmarks in a continuous two-dimensional environment. The maximum episode length is 25, i.e. episodes are terminated after 25 steps. All agents receive their velocity, position, and relative position to all other agents and landmarks.  
% The action space of each agent is five discrete movement actions. The reward of each agent include two part, the first part is sum of negative minimum distances from each landmark to any agent, the second part is the punish to collisions among agents.

\noindent \textbf{Battle} 
The \emph{Battle} scenario has two teams including Red and Blue, each team has 121 agents. The map size is $80\times80$. Agents slowly regain HP, so it is best to kill an opposing agent quickly. Agents have 10 HP at the beginning, minus 2 HP by each attack, and recover 0.1 HP every step. Each agent will be rewarded 0.2 when they attack the opponent agent and be punished -0.1 when dead. Also, they will get a -0.1 penalty for each attack action. 
% Therefore, the goal of each agent is to reduce its own damage while destroying as many opponents as possible. The maximum episode length is 200. 

\noindent \textbf{Battlefiled}  
The settings for \emph{Battle} and \emph{Battlefiled} are basically the same, but the \emph{Battlefiled} has some fixed obstacles which make the task become more difficult. The map size is $80\times 80$, and the agent in each team is 74.

\noindent \textbf{Gather} 
The \emph{Gather} has hundreds of agents and thousands of pieces of food in this environment.  The agent will gain reward by eating food. Each food will be absorbed after 5 attacks. Since the food is limited, so the agents may compete with each other to monopolize food. 

\subsection{Results on Cooperative Navigation}
There are N agents and L landmarks in the \emph{Cooperative Navigation}. Each agent can observe the relative positions of other agents and landmarks, collectively rewarded based on the distance of any agent to each landmark, and are penalized when colliding with each other. So each agent should learn to occupy more landmarks with fewer collisions. However, the origin setting $N = 3, L = 3$ is too simple to distinguish the performance between different models. Therefore we set the $N = 6, L = 3$, the agent needs to learn how to share the landmark with other agents.

To answer \textbf{Q1}, we combine SEA with different classical algorithms, including A2C, PPO, and DDPG. The results are shown in Fig.\ref{fig:mpe_curve}. We carry out each experiment with five random seeds, the shaded area is enclosed by the min and max value of all random seeds, and the solid line is the mean value of all seeds (same for all experiments). 
% We use the max operation (as in Eq.\ref{pointnet}) for each cluster to group features in the main results and the mean operation (as in Eq.\ref{avg}) experiment can be found in Appendix B. 
The results show that by adding a spatial extract component without any other change, all algorithms significantly improved compared to their original algorithm. As shown in Fig.\ref{a2c_}, A2C is quite unstable that several seeds have a massive fall during the training period; MAA2C has the global state in critic, however, it does not go beyond A2C, it may be due to the global state in MPE is simply concatenate observation of all agents together which will cause an abundance of redundant information. SEA-A2C is more stable, and the performance is beyond the A2C and MAA2C, which means that SEA can give better information sharing between agents. 
As shown in Fig.\ref{ppo_}, similar to the situation in A2C, but MAPPO performs better than MAA2C, the results of MAPPO are close to PPO and is more stable, the final results show that SEA-PPO achieves significant advantages over the PPO and MAPPO. Different from the two on-policy methods above, in Fig.\ref{ddpg_}, DDPG performs worse than MADDPG, and the SEA-DDPG also improves a lot compared to the MADDPG and DDPG. Since the SEA-DDPG has information sharing between agents, for a fair comparison, we also evaluate the MAAC, which uses attention to sharing information between agents, and MAAC is also an off-policy method. The results in Fig.\ref{ddpg_} show that MAAC can achieve a high score faster, but it is still not beyond the SEA-DDPG, which can answer the \textbf{Q2}.
% and parameter setting 

% \subsection{Results on Predator Prey}

% \begin{figure}[h]
% \subfigure[Battle]{
% \begin{minipage}{0.48\linewidth}
% \centerline{\includegraphics[width=1.1\textwidth]{figure/battle_121_train_reward.pdf}}
% % \centerline{}
% \label{fig:battle}
% \end{minipage}}
% \subfigure[Battle Filed]{
% \begin{minipage}{0.48\linewidth}
% \centerline{\includegraphics[width=1.1\textwidth]{figure/gather_495_train_reward.pdf}}
% % \centerline{}
% \label{fig:gather}
% \end{minipage}}
% \caption{Main results on Battle and Battle Filed environments.}
% \label{fig:magent}
% \end{figure}

\subsection{Results on Battle and Battlefiled}
In the \emph{Battle}, two teams need to battle with the opponent team, and each team has $N$ agents. The scale of each team can be set to different sizes. Each agent can move or attack at each time step and will be rewarded for their performance. Specifically, each agent will be rewarded when attacking the opponent agents and punished when dead. 

We set the pretrained DQN model as the fixed opponent strategy and trained all the models with 64 agents setting for 2000 episodes. The main results in Battle are shown in Fig.\ref{fig:battle}. SEA-A2C converges faster than all other baselines; MFQ and DQN perform similarly. Although MFAC learns faster at the beginning, its training curve is unstable, and the final reward is lower than other methods. Since the Battle scenario is quite simple, the differences between algorithms are small.

The \emph{Battlefield} is more complex than \emph{Battle} because there are some obstacles in Battlefield, which makes agent learning more difficult. Since the map is bigger, we set 74 agents as our main experiment setting. The results in Fig.\ref{fig:battlefiled} shows that in the more complex scenario, the SEA-A2C can have a faster convergence speed than all other methods, and it is more stable than other methods. DQN and MFQ first get a relatively high score. However, since those agents who are more aggressive or in front of the group will die first, their sample scale will be less, so the model will be easier to learn a passive strategy, that is, stay still or avoid, leading to convergence to a much lower reward. The MFAC performs well initially, but the training process is unstable and converges slower.

\subsection{Results on Gather}
In \emph{Gather}, there are hundreds of agents and thousands of pieces of food. The agents gain rewards by eating food. Food needs to break down by five attacks before it is absorbed. The number of food is limited, so the agents could cooperate with other agents to obtain food more effectively or attack others to monopolize the food.

\begin{figure*}[ht]
    \centering
    \includegraphics[width=6.2 in]{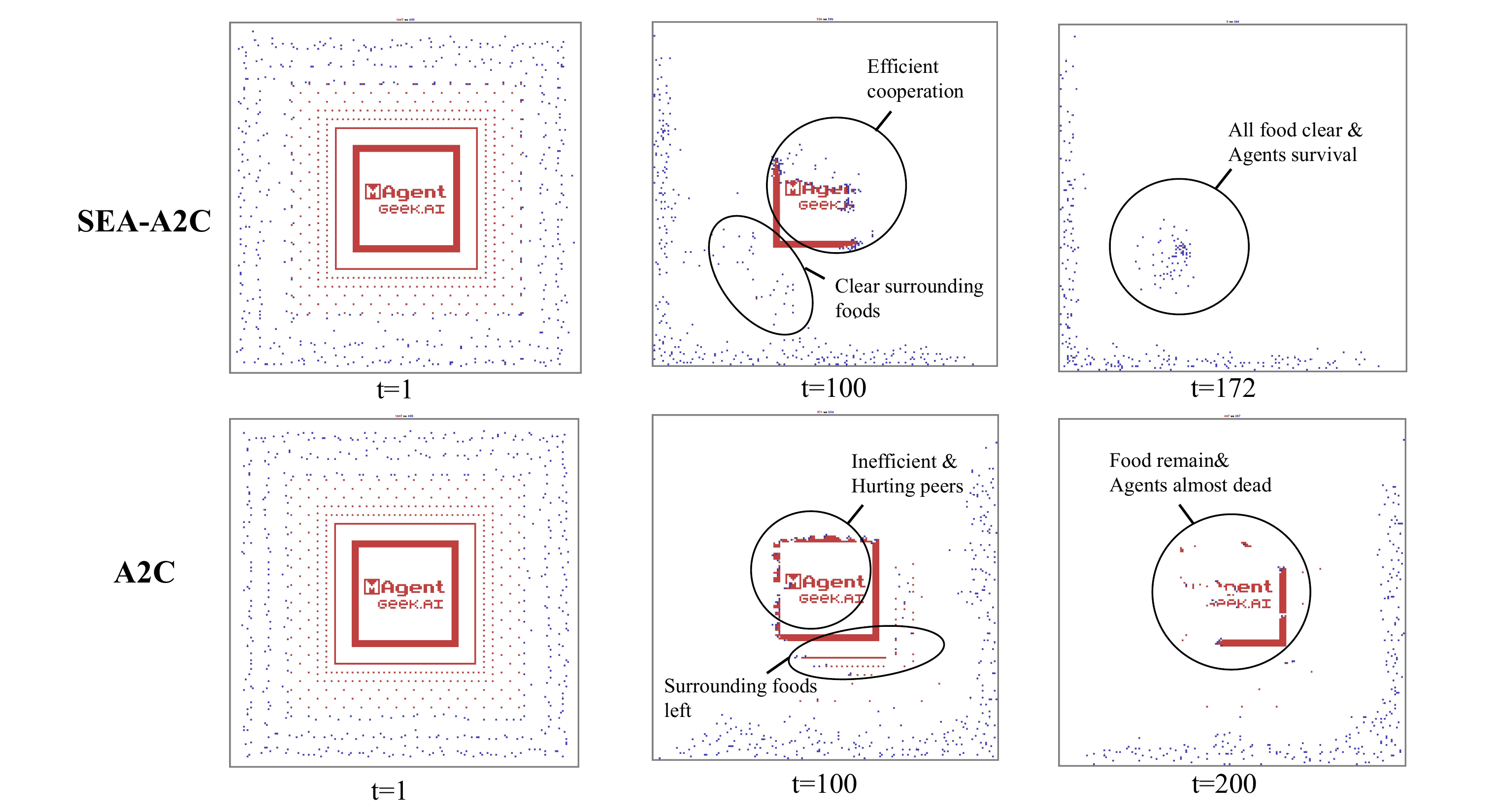}
    \caption{The analysis and visualization of SEA-A2C and A2C in Gather scenario. The max limit of the time step is 200. From left to right is the timeline of each algorithm from the start, middle, and end of the episode.}
    \label{fig:vis}
\end{figure*}

\paragraph{Main results}
The main results in \emph{Gather} are shown in Fig.\ref{fig:gather}. Here we use the default setting of 495 agents and 1847 foods, but we limit the number of time steps here to let agents can concern more with absorbing food. SEA-AC converges to a much higher reward than other baselines and achieves a more stable training curve. 

To further investigate the reason why SEA-A2C can achieve such an advantage on reward, we evaluate SEA-A2C and other baselines by running 10 test games, and each game is unrolled with 200 steps. The detailed information, including total reward, dead count, food absorbed, and the absorbed-deaths ratio of each algorithm shown in Table~\ref{tab:gather}. For those independent methods (such as DQN and A2C), they try to get as much food as possible and kill other agents to monopolize the food or just become more passive than not competing with the food. Therefore, there is a trade-off between agent deaths and food absorbed. The SEA-A2C gets the highest absorbed-death ratio and absorbs almost all food, proving that the SEA-A2C can effectively balance this trade-off. A2C and MFAC can absorb about 70\% food in a limited time, but more agents need to be sacrificed; MFQ and DQN fall into another extreme that is too afraid of killing that only absorb little food.
\begin{table}[htbp]
    \centering
    \caption{The Details of Gather Results}
    \begin{tabular}{lllll}
        \hline
        Algorithms  & Reward & Absorbed & Deaths & Ratio\\
        \hline
        SEA-A2C     & \textbf{10911}        & \textbf{1845}(99\%)  &  217 & \textbf{8.50} \\
        A2C   & 7679         &  1302(70\%)   & 272   & 4.78\\
        MFAC   &  7150      &  1361(73\%)  &  288    & 4.72\\
        MFQ & 3101         & 538(29\%)   & \textbf{86} & 6.25\\
        DQN  & 766      &   364(20 \%)  &  97  & 3.75\\
        \hline
    \end{tabular}
    \label{tab:gather}
\end{table}

% \subsubsection{Generalization.}
% - training on number x, test on number y.

\begin{figure}[b]
    \centering
    \includegraphics[width=3 in]{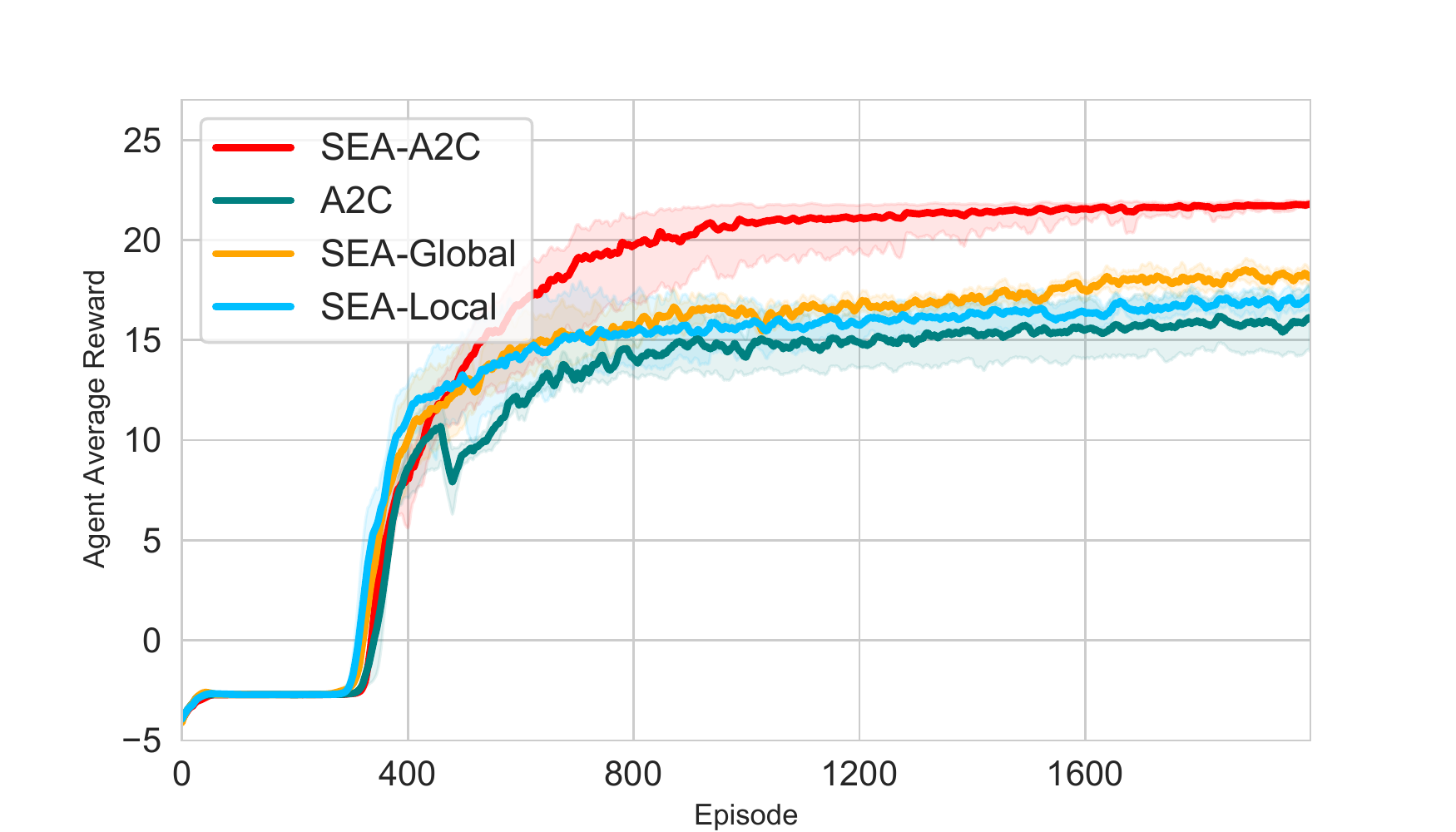}
    \caption{Ablation study on gather.}
    \label{fig:ablation}
\end{figure}
\paragraph{Case Study and Visualization}
In order to show the specific sources of SEA-A2C's advantages in experimental results, we visualized the process of test results in Fig.\ref{fig:vis}. As observed in the experiment, under the same scale as the beginning setting, because agents share global and local information while learning, SEA-A2C is more inclined to cooperate with others than kill others. When t=100, a large number of agents of SEA-A2C perform efficient cooperative mining in the center, and the peripheral agents clean up the scattered food at the corner. The agents of the A2C algorithm mining food in the center of the map kill each other because they want to monopolize food, resulting in low efficiency. The SEA-A2C algorithm completed all food collection tasks in advance in step 172, but the A2C algorithm did not complete food collection until step 200, and most agents in the middle died from killing each other. Since we follow the CTDE settings, that each agent does not share info during execution and only has local observation, those agents on edge have no food information in their vision, so no matter SEA-A2C or A2C algorithm, they can only wander on edge.

\paragraph{Ablations}
Since SEA can extract both local and global information between agents, we do an ablation study in this section to investigate further how the local and global information affects the performance of SEA (\textbf{Q3}). We designed two variants to extract only local or global information. In the SEA-Global, we group all points into one cluster, therefore the one cluster has the aggregate information of all points. In the SEA-Local, we remove the high level grouping so that SEA will only share information between nearby agents. The results in Fig.\ref{fig:ablation} show that both global and local information can improve A2C, but the degree is minimal. The model can be greatly improved when global and local features are obtained simultaneously.

% - Avg and Max

% - group all and only local

% - mask or not

% - sight vision

% \subsubsection{Cost Analyse}
% - time cost with the increase of agent scale, compare with DGN/mfac/ac.

\section{Conclusion \& Future Work}
This paper proposes a spatially explicit framework under the CTDE paradigm for MARL called SEA. We combine SEA with various multi-agent algorithms and evaluate their performances in several environments. The results show that SEA can effectively share local and global hidden information between agents during training, especially when the scale of agents is large and the task is complex. To the best of our knowledge, SEA is the first study that combines the point cloud problem in computer vision with the multi-agent decision-making problem. In the future, we plan to discuss further the performance of SEA in non-CTDE settings.

\bibliographystyle{IEEEtran}
\bibliography{IEEEabrv,./ref}

\begin{thebibliography}{10}
\providecommand{\url}[1]{#1}
\csname url@samestyle\endcsname
\providecommand{\newblock}{\relax}
\providecommand{\bibinfo}[2]{#2}
\providecommand{\BIBentrySTDinterwordspacing}{\spaceskip=0pt\relax}
\providecommand{\BIBentryALTinterwordstretchfactor}{4}
\providecommand{\BIBentryALTinterwordspacing}{\spaceskip=\fontdimen2\font plus
\BIBentryALTinterwordstretchfactor\fontdimen3\font minus
  \fontdimen4\font\relax}
\providecommand{\BIBforeignlanguage}[2]{{%
\expandafter\ifx\csname l@#1\endcsname\relax
\typeout{** WARNING: IEEEtran.bst: No hyphenation pattern has been}%
\typeout{** loaded for the language `#1'. Using the pattern for}%
\typeout{** the default language instead.}%
\else
\language=\csname l@#1\endcsname
\fi
#2}}
\providecommand{\BIBdecl}{\relax}
\BIBdecl

\bibitem{dota2drl}
C.~Berner, G.~Brockman, B.~Chan, V.~Cheung, P.~Debiak, C.~Dennison, D.~Farhi,
  Q.~Fischer, S.~Hashme, C.~Hesse, R.~J{\'{o}}zefowicz, S.~Gray, C.~Olsson,
  J.~Pachocki, M.~Petrov, H.~P. de~Oliveira~Pinto, J.~Raiman, T.~Salimans,
  J.~Schlatter, J.~Schneider, S.~Sidor, I.~Sutskever, J.~Tang, F.~Wolski, and
  S.~Zhang, ``Dota 2 with large scale deep reinforcement learning,''
  \emph{CoRR}, vol. abs/1912.06680, 2019.

\bibitem{starcraft2}
O.~Vinyals, I.~Babuschkin, W.~M. Czarnecki, M.~Mathieu, A.~Dudzik, J.~Chung,
  D.~H. Choi, R.~Powell, T.~Ewalds, P.~Georgiev, J.~Oh, D.~Horgan, M.~Kroiss,
  I.~Danihelka, A.~Huang, L.~Sifre, T.~Cai, J.~P. Agapiou, M.~Jaderberg, A.~S.
  Vezhnevets, R.~Leblond, T.~Pohlen, V.~Dalibard, D.~Budden, Y.~Sulsky,
  J.~Molloy, T.~L. Paine, {\c{C}}.~G{\"{u}}l{\c{c}}ehre, Z.~Wang, T.~Pfaff,
  Y.~Wu, R.~Ring, D.~Yogatama, D.~W{\"{u}}nsch, K.~McKinney, O.~Smith,
  T.~Schaul, T.~P. Lillicrap, K.~Kavukcuoglu, D.~Hassabis, C.~Apps, and
  D.~Silver, ``Grandmaster level in starcraft {II} using multi-agent
  reinforcement learning,'' \emph{Nat.}, vol. 575, no. 7782, pp. 350--354,
  2019.

\bibitem{trafic_chu}
T.~Chu, J.~Wang, L.~Codec{\`{a}}, and Z.~Li, ``Multi-agent deep reinforcement
  learning for large-scale traffic signal control,'' \emph{{IEEE} Trans.
  Intell. Transp. Syst.}, vol.~21, no.~3, pp. 1086--1095, 2020.

\bibitem{autonomous_driving}
S.~Shalev{-}Shwartz, S.~Shammah, and A.~Shashua, ``Safe, multi-agent,
  reinforcement learning for autonomous driving,'' \emph{CoRR}, vol.
  abs/1610.03295, 2016.

\bibitem{independent_learning}
M.~Lanctot, V.~F. Zambaldi, A.~Gruslys, A.~Lazaridou, K.~Tuyls,
  J.~P{\'{e}}rolat, D.~Silver, and T.~Graepel, ``A unified game-theoretic
  approach to multiagent reinforcement learning,'' in \emph{Advances in Neural
  Information Processing Systems 30: Annual Conference on Neural Information
  Processing Systems 2017, December 4-9, 2017, Long Beach, CA, {USA}},
  I.~Guyon, U.~von Luxburg, S.~Bengio, H.~M. Wallach, R.~Fergus, S.~V.~N.
  Vishwanathan, and R.~Garnett, Eds., 2017, pp. 4190--4203.

\bibitem{Busoniu2010}
L.~Bu{\c{s}}oniu, R.~Babu{\v{s}}ka, and B.~De~Schutter, \emph{Multi-agent
  Reinforcement Learning: An Overview}.\hskip 1em plus 0.5em minus 0.4em\relax
  Berlin, Heidelberg: Springer Berlin Heidelberg, 2010, pp. 183--221.

\bibitem{mfrl}
Y.~Yang, R.~Luo, M.~Li, M.~Zhou, W.~Zhang, and J.~Wang, ``Mean field
  multi-agent reinforcement learning,'' in \emph{Proceedings of the 35th
  International Conference on Machine Learning, {ICML} 2018,
  Stockholmsm{\"{a}}ssan, Stockholm, Sweden, July 10-15, 2018}, ser.
  Proceedings of Machine Learning Research, J.~G. Dy and A.~Krause, Eds.,
  vol.~80.\hskip 1em plus 0.5em minus 0.4em\relax {PMLR}, 2018, pp. 5567--5576.

\bibitem{ncc}
H.~Mao, W.~Liu, J.~Hao, J.~Luo, D.~Li, Z.~Zhang, J.~Wang, and Z.~Xiao,
  ``Neighborhood cognition consistent multi-agent reinforcement learning,'' in
  \emph{The Thirty-Fourth {AAAI} Conference on Artificial Intelligence, {AAAI}
  2020, The Thirty-Second Innovative Applications of Artificial Intelligence
  Conference, {IAAI} 2020, The Tenth {AAAI} Symposium on Educational Advances
  in Artificial Intelligence, {EAAI} 2020, New York, NY, USA, February 7-12,
  2020}.\hskip 1em plus 0.5em minus 0.4em\relax {AAAI} Press, 2020, pp.
  7219--7226.

\bibitem{mfvfd}
T.~Zhang, Q.~Ye, J.~Bian, G.~Xie, and T.~Liu, ``{MFVFD:} {A} multi-agent
  q-learning approach to cooperative and non-cooperative tasks,'' in
  \emph{Proceedings of the Thirtieth International Joint Conference on
  Artificial Intelligence, {IJCAI} 2021, Virtual Event / Montreal, Canada,
  19-27 August 2021}, Z.~Zhou, Ed.\hskip 1em plus 0.5em minus 0.4em\relax
  ijcai.org, 2021, pp. 500--506.

\bibitem{DGN}
J.~Jiang, C.~Dun, and Z.~Lu, ``Graph convolutional reinforcement learning for
  multi-agent cooperation,'' \emph{CoRR}, vol. abs/1810.09202, 2018.

\bibitem{commNet}
S.~Sukhbaatar, A.~Szlam, and R.~Fergus, ``Learning multiagent communication
  with backpropagation,'' in \emph{Advances in Neural Information Processing
  Systems 29: Annual Conference on Neural Information Processing Systems 2016,
  December 5-10, 2016, Barcelona, Spain}, D.~D. Lee, M.~Sugiyama, U.~von
  Luxburg, I.~Guyon, and R.~Garnett, Eds., 2016, pp. 2244--2252.

\bibitem{bicnet}
P.~Peng, Q.~Yuan, Y.~Wen, Y.~Yang, Z.~Tang, H.~Long, and J.~Wang, ``Multiagent
  bidirectionally-coordinated nets for learning to play starcraft combat
  games,'' \emph{CoRR}, vol. abs/1703.10069, 2017.

\bibitem{gridnet}
L.~Han, P.~Sun, Y.~Du, J.~Xiong, Q.~Wang, X.~Sun, H.~Liu, and T.~Zhang,
  ``Grid-wise control for multi-agent reinforcement learning in video game
  {AI},'' in \emph{Proceedings of the 36th International Conference on Machine
  Learning, {ICML} 2019, 9-15 June 2019, Long Beach, California, {USA}}, ser.
  Proceedings of Machine Learning Research, K.~Chaudhuri and R.~Salakhutdinov,
  Eds., vol.~97.\hskip 1em plus 0.5em minus 0.4em\relax {PMLR}, 2019, pp.
  2576--2585.

\bibitem{pointnet}
C.~R. Qi, H.~Su, K.~Mo, and L.~J. Guibas, ``Pointnet: Deep learning on point
  sets for 3d classification and segmentation,'' \emph{CoRR}, vol.
  abs/1612.00593, 2016.

\bibitem{pointnet2}
C.~R. Qi, L.~Yi, H.~Su, and L.~J. Guibas, ``Pointnet++: Deep hierarchical
  feature learning on point sets in a metric space,'' in \emph{Advances in
  Neural Information Processing Systems 30: Annual Conference on Neural
  Information Processing Systems 2017, December 4-9, 2017, Long Beach, CA,
  {USA}}, I.~Guyon, U.~von Luxburg, S.~Bengio, H.~M. Wallach, R.~Fergus,
  S.~V.~N. Vishwanathan, and R.~Garnett, Eds., 2017, pp. 5099--5108.

\bibitem{maddpg}
R.~Lowe, Y.~Wu, A.~Tamar, J.~Harb, P.~Abbeel, and I.~Mordatch, ``Multi-agent
  actor-critic for mixed cooperative-competitive environments,'' in
  \emph{Advances in Neural Information Processing Systems 30: Annual Conference
  on Neural Information Processing Systems 2017, December 4-9, 2017, Long
  Beach, CA, {USA}}, I.~Guyon, U.~von Luxburg, S.~Bengio, H.~M. Wallach,
  R.~Fergus, S.~V.~N. Vishwanathan, and R.~Garnett, Eds., 2017, pp. 6379--6390.

\bibitem{coma}
J.~N. Foerster, G.~Farquhar, T.~Afouras, N.~Nardelli, and S.~Whiteson,
  ``Counterfactual multi-agent policy gradients,'' in \emph{Proceedings of the
  Thirty-Second {AAAI} Conference on Artificial Intelligence, (AAAI-18), the
  30th innovative Applications of Artificial Intelligence (IAAI-18), and the
  8th {AAAI} Symposium on Educational Advances in Artificial Intelligence
  (EAAI-18), New Orleans, Louisiana, USA, February 2-7, 2018}, S.~A. McIlraith
  and K.~Q. Weinberger, Eds.\hskip 1em plus 0.5em minus 0.4em\relax {AAAI}
  Press, 2018, pp. 2974--2982.

\bibitem{mappo}
C.~Yu, A.~Velu, E.~Vinitsky, Y.~Wang, A.~M. Bayen, and Y.~Wu, ``The surprising
  effectiveness of {MAPPO} in cooperative, multi-agent games,'' \emph{CoRR},
  vol. abs/2103.01955, 2021.

\bibitem{efficient}
\BIBentryALTinterwordspacing
B.~Zhang, Y.~Bai, Z.~Xu, D.~Li, and G.~Fan, ``Efficient cooperation strategy
  generation in multi-agent video games via hypergraph neural network,''
  \emph{CoRR}, vol. abs/2203.03265, 2022. [Online]. Available:
  \url{https://doi.org/10.48550/arXiv.2203.03265}
\BIBentrySTDinterwordspacing

\bibitem{vdn}
P.~Sunehag, G.~Lever, A.~Gruslys, W.~M. Czarnecki, V.~F. Zambaldi,
  M.~Jaderberg, M.~Lanctot, N.~Sonnerat, J.~Z. Leibo, K.~Tuyls, and T.~Graepel,
  ``Value-decomposition networks for cooperative multi-agent learning,''
  \emph{CoRR}, vol. abs/1706.05296, 2017.

\bibitem{qmix}
T.~Rashid, M.~Samvelyan, C.~S. de~Witt, G.~Farquhar, J.~N. Foerster, and
  S.~Whiteson, ``{QMIX:} monotonic value function factorisation for deep
  multi-agent reinforcement learning,'' in \emph{Proceedings of the 35th
  International Conference on Machine Learning, {ICML} 2018,
  Stockholmsm{\"{a}}ssan, Stockholm, Sweden, July 10-15, 2018}, ser.
  Proceedings of Machine Learning Research, J.~G. Dy and A.~Krause, Eds.,
  vol.~80.\hskip 1em plus 0.5em minus 0.4em\relax {PMLR}, 2018, pp. 4292--4301.

\bibitem{side}
\BIBentryALTinterwordspacing
Z.~Xu, Y.~Bai, D.~Li, B.~Zhang, and G.~Fan, ``{SIDE:} {I} infer the state {I}
  want to learn,'' \emph{CoRR}, vol. abs/2105.06228, 2021. [Online]. Available:
  \url{https://arxiv.org/abs/2105.06228}
\BIBentrySTDinterwordspacing

\bibitem{haven}
\BIBentryALTinterwordspacing
Z.~Xu, Y.~Bai, B.~Zhang, D.~Li, and G.~Fan, ``{HAVEN:} hierarchical cooperative
  multi-agent reinforcement learning with dual coordination mechanism,''
  \emph{CoRR}, vol. abs/2110.07246, 2021. [Online]. Available:
  \url{https://arxiv.org/abs/2110.07246}
\BIBentrySTDinterwordspacing

\bibitem{mbvd}
\BIBentryALTinterwordspacing
Z.~Xu, D.~Li, B.~Zhang, Y.~Zhan, Y.~Bai, and G.~Fan, ``Mingling foresight with
  imagination: Model-based cooperative multi-agent reinforcement learning,''
  \emph{CoRR}, vol. abs/2204.09418, 2022. [Online]. Available:
  \url{https://doi.org/10.48550/arXiv.2204.09418}
\BIBentrySTDinterwordspacing

\bibitem{serl}
B.~Yan, K.~Janowicz, G.~Mai, and R.~Zhu, ``A spatially explicit reinforcement
  learning model for geographic knowledge graph summarization,'' \emph{Trans.
  {GIS}}, vol.~23, no.~3, pp. 620--640, 2019.

\bibitem{pool}
Z.~Cao, X.~Ma, M.~Shi, and Z.~Zhao, ``Pheromone-inspired communication
  framework for large-scale multi-agent reinforcement learning,'' in
  \emph{Artificial Neural Networks and Machine Learning - {ICANN} 2022 - 31st
  International Conference on Artificial Neural Networks, Bristol, UK,
  September 6-9, 2022, Proceedings, Part {II}}, ser. Lecture Notes in Computer
  Science, E.~Pimenidis, P.~P. Angelov, C.~Jayne, A.~Papaleonidas, and
  M.~Aydin, Eds., vol. 13530.\hskip 1em plus 0.5em minus 0.4em\relax Springer,
  2022, pp. 75--86.

\bibitem{magent}
L.~Zheng, J.~Yang, H.~Cai, M.~Zhou, W.~Zhang, J.~Wang, and Y.~Yu, ``Magent: {A}
  many-agent reinforcement learning platform for artificial collective
  intelligence,'' in \emph{Proceedings of the Thirty-Second {AAAI} Conference
  on Artificial Intelligence, (AAAI-18), the 30th innovative Applications of
  Artificial Intelligence (IAAI-18), and the 8th {AAAI} Symposium on
  Educational Advances in Artificial Intelligence (EAAI-18), New Orleans,
  Louisiana, USA, February 2-7, 2018}, S.~A. McIlraith and K.~Q. Weinberger,
  Eds.\hskip 1em plus 0.5em minus 0.4em\relax {AAAI} Press, 2018, pp.
  8222--8223.

\bibitem{lowe2017multi}
R.~Lowe, Y.~Wu, A.~Tamar, J.~Harb, P.~Abbeel, and I.~Mordatch, ``Multi-agent
  actor-critic for mixed cooperative-competitive environments,'' \emph{Neural
  Information Processing Systems (NIPS)}, 2017.

\end{thebibliography}
\end{document}